\documentclass{aa}
\usepackage{epsfig}
\usepackage{txfonts}

\usepackage{epsfig}
\newcommand{\msun}{\mbox{$M_{\odot}$}}

\newcommand{\logl}{\mbox{$\log (L/L_{\odot}$)}}  

\newcommand{\zsun}{\mbox{$Z_{\odot}$}}
\newcommand{\teff}{\mbox{$T_{\rm eff}$}}

\newcommand{\vinf}{\mbox{$v_{\infty}$}}

\newcommand{\mdot}{\mbox{$\dot{M}$}}

\newcommand{\tstar}{\mbox{$T_{*}$}}
\newcommand{\msunyr}{\mbox{$M_{\odot} {\rm yr}^{-1}$}}

\newcommand{\rstar}{\mbox{$R_{*}$}}   
\newcommand{\reff}{\mbox{$R_{\rm eff}$}}
\newcommand{\kms}{km s$^{-1}$}

\begin{document}


\title{On the metallicity dependence of Wolf-Rayet winds}

\author{Jorick S. Vink\inst{1}
\and Alex de Koter\inst{2}}
\offprints{Jorick S. Vink, jsv@astro.keele.ac.uk}

\institute{Imperial College, Blackett Laboratory, 
           Prince Consort Road, London, SW7 2AZ, U.K.
            \and
            Astronomical Institute ``Anton Pannekoek'', University of Amsterdam,
           Kruislaan 403, NL-1098 SJ Amsterdam, The Netherlands}

\titlerunning{Mass loss versus $Z$ for WR stars}
\authorrunning{Jorick S. Vink \& Alex de Koter}

\abstract{We have performed a pilot study of mass loss predictions
for late-type Wolf-Rayet (WR) stars as a function of metal
abundance, over a range between $10^{-5} \le ($Z$/\zsun) \le 10$.  We
find that the winds of nitrogen-rich Wolf-Rayet stars are
dominated by iron lines, with a dependence of mass loss on $Z$
similar to that of massive OB stars. For more evolved, carbon-rich, WR
stars the wind strength is also found to be dependent on the Fe abundance,
so that they depend on the chemical environment of the host galaxy, but with a  
mass loss metallicity dependence that is less steep than for OB stars.  
Our finding that WR mass loss is $Z$-dependent is a new one, with 
important consequences for black hole formation and X-ray population
studies in external galaxies.
A further finding of our study is that the $Z$ dependence of
C-rich WR stars becomes weaker
at metallicities below $Z/\zsun$ $\la$ 1/10, and {\it mass loss 
no longer declines once the metal abundance drops} below ($Z$/\zsun) $\simeq$ $10^{-3}$. 
This is the result of an increased importance of
radiative driving by intermediate mass elements, such as carbon. 
In combination with rapid rotation and/or proximity to the 
Eddington limit -- likely to be relevant for massive Population {\sc iii} stars -- this effect 
may indicate a role for mass loss in the appearance and evolution of these objects, as well as a potential role 
for stellar winds in enriching the intergalactic medium of the early Universe.
}

\maketitle


\section{Introduction}
\label{s_intro}

The physics of mass loss from massive stars is fundamental for a range
of astrophysical issues, comprising the formation and evolution of
individual massive stars, their final fate, in particular the nature of 
the compact object that is left behind (neutron star or black hole), 
as well as the input of energy, ionizing radiation, and chemical
elements into the interstellar medium (ISM) of the host galaxy.
There is growing evidence that long-duration gamma-ray bursts (GRBs) are the result
of the collapse of a massive star (Galama et al. 1998, Hjorth et
al. 2003).  As these cosmological explosions are seen up to high
redshift, and consequently at very low metal abundances, one
needs to establish the $Z$-dependent properties of their progenitors,
quite likely WR stars (e.g. Woosley et al. 1999). 
Furthermore, the observed properties of GRB are expected
to depend on the history of the wind strength of the progenitor.
The question that therefore needs to be addressed is: ``are the winds of
WR stars metallicity dependent?''

Although there are indirect indications for such a 
metallicity dependence (e.g. Hadfield et al. 2005), a theoretical prediction 
for metallicity dependent WR winds is currently unavailable, nor is there a 
well-established observational relationship for it. Consequently, stellar 
evolution theorists need to opt for one of two rather drastic assumptions in dealing
with this issue: either, ({\it i)} 
they extrapolate the mass loss - $Z$ dependence of O stars to WR stars (e.g. Heger \& Woosley 2002, Heger et al. 2003, 
Eldridge \& Tout 2004), or ({\it ii)} they
assume a mass loss that is independent of $Z$ at low metallicity (e.g. Schaller et
al. 1992, Maeder \& Meynet 1994, Meynet \& Maeder 2005).

The first assumption is based on the idea that O and WR stars share
similar properties: both are hot, luminous, and generate high-speed outflows. Therefore, 
one might expect the winds to be driven by the same physical
mechanism, i.e. that of radiation pressure on millions of spectral
lines (Lucy \& Solomon 1970; Castor et al. 1975, hereafter CAK). The
second assumption arises from the fact that WR stars produce 
copious amounts of carbon and oxygen through self-enrichment, so that the radiation-driven wind may 
not depend on the initial metallicity of the host galaxy.

That WR winds could at all 
be driven by radiation pressure on spectral lines (as O star winds), was in doubt 
for many decades,
as observations of WR stars showed particularly high wind momenta with 
values of the product of the mass-loss rate and the terminal flow
velocity, $\mdot \vinf$, 10 -- 100 times larger than that of the
momentum of the stellar radiation field, $L_{*}/c$ -- the
so-called ``wind-momentum problem'' (e.g. Barlow et al. 1981), which in
effect was an opacity problem (Gayley et al. 1995).
Developments over the last years have shown that radiation pressure may be a viable 
mechanism for both initiating as well as 
driving the winds of WR stars after all.

One reason is the discovery
that WR stars are highly clumped (e.g. Moffat \& Robert 1994, Lepine
\& Moffat 1999), which has resulted in a downward revision of the
observed WR mass-loss rates (e.g. Hamann \& Koesterke 1998).
A second reason is the realization that the ionization structure
of WR winds is stratified: at progressively deeper layers, higher
ionization stages become dominant, which increases the number of lines 
available for wind driving, whilst simultaneously 
line photons are allowed to be scattered more than once -- increasing 
their effectiveness in transferring momentum to the gas (Lucy \& Abbott
1993, Springmann 1994, Gayley 1995, de Koter et al. 1997).  
This combined effect of ionization stratification and multiple
line scattering leads to predictions for the ratio of wind momentum
over radiative momentum , $\eta = \mdot \vinf / (L/c)$, of up to a
factor of 10 -- alleviating the wind momentum problem (but see Brown et al. 2004).

Last but not least, it appears that the ``hot iron bump'' (Fe {\sc x} -- Fe {\sc xiv})
in the deep layers of WR stars has been identified as the initiation
region of WR winds, 
conceptually proposed by Nugis \& Lamers (2002), and 
most recently implemented numerically by
Gr\"{a}fener \& Hamann (2005).  These latter authors could self-consistently 
solve the dynamics for a carbon-rich
WC5 star, thereby showing that radiation pressure is capable of
initiating and maintaining the entire wind from below the photosphere
to the region where the wind reaches its terminal velocity. 

Now that it appears that radiation pressure is the principal driving
mechanism for WR winds, we believe the time is ripe to explore the
metallicity dependence of WR winds, as has previously been done 
for O stars (Vink et al. 2001, Kudritzki 2002).  Empirically, 
the issue of an \mdot\ ($Z$)-dependence for WR stars has been
studied by Crowther et al. (2002). Comparing WR stars in the Galaxy 
and the Large Magellanic Cloud, they report a tentative 
\mdot(Z) dependence, however this result is not clear-cut, in part because
the investigated baseline in $Z$ is rather narrow (about a factor of two in $Z$). 

As WR stars produce copious amounts of carbon and oxygen, one might
argue (referring back to the second assumption mentioned earlier) 
that WR mass loss is likely to be {\it independent} of the
initial chemical composition (which is set by the metallicity of the host galaxy),
once intermediate mass elements become abundant in the outer layers. 
The relevant question is therefore whether
the lines of intermediate elements do indeed dominate the Fe lines 
in their contribution to the total line force, and if so, whether 
this holds true for all metallicities, i.e. Fe mass fractions. Or 
alternatively, whether the Fe lines may take over the line driving 
starting at some minimum value of $Z$. 

For O stars, the relative importance for the wind driving of
intermediate mass elements (mainly CNO) versus that
of heavy metals (notably Fe) as a function of $Z$ has been
studied in depth by Vink et al. (2001). 
These authors found that the intermediate elements of CNO, and to some
extent even hydrogen and helium, gain importance in driving the winds
of O stars at $Z$ $\la$ 1/30 giving rise to a flattening of the O star
mass-loss rate versus metallicity relation (see their Fig~5c). 
Interestingly, the opposite behaviour has been found by Kudritzki (2002),
who finds that $\mdot$ decreases even more rapidly at $Z/\zsun$ $\la$
$10^{-3}$.
The cause of this discrepancy at very low $Z$ in the $\mdot$(Z) behaviour between 
the two approaches has yet to be identified. It may be related to their treatments 
of the relevant aspects of the physics of line driving at different levels of sophistication. 
Kudritzki (2002) parameterizes the radiative force through so-called force
multipliers (in an extension of CAK theory), enabling a relatively straightforward   
solution of the wind dynamics.
Vink et al. (2001) calculate the radiative force exactly with a 
Monte Carlo approach that allows for a detailed treatment of multiple 
photon scatterings, albeit adopting 
a pre-described velocity law (but see Vink et al. 1999, Vink 2000). The global approach 
works very well for OB stars (Vink et al. 2000), and although its validity for WR stars 
is less well-established, we use it here, noting
that the $\beta$-law model of Gr\"{a}fener \& Hamann (2005) shows a similar 
global energy budget as their self-consistent computation for the same parameters. 
This indicates that the global method may 
produce meaningful results (as for O stars). 
In any case, it is our philosophy to accept small errors in the 
mass-loss determination, as our goal is to explore a large
parameter space, comprising a wide range of metallicities and mass-loss rates, 
and to establish the $\mdot(Z)$ scaling law that may already be retrieved from a 
differential study.

In the next two sections, we describe our method and assumptions.  Our basic 
Wolf-Rayet wind models are presented in Sect.~\ref{s_massloss}, followed by a 
study of their mass-loss properties as a function of metal abundance. 
We make a distinction
between the nitrogen-rich Wolf-Rayet stars (WN), which may still
contain some hydrogen, but with helium present in large quantities,
and the carbon rich WR stars (WC) that have converted a
significant fraction of helium into carbon, thereby representing a
later evolutionary phase.  We discuss the implications of our findings
in Sect.~\ref{s_impl}, and summarize in Sect.~\ref{s_concl}

\section{Method to calculate \mdot\ with Monte Carlo}
\label{s_method}

The mass-loss rates are calculated with a Monte Carlo (MC) code that is used 
to follow the fate of a large number of photons from below the stellar
photosphere through the wind, and that calculates the radiative
acceleration of the wind material. The core of the approach is that  
the cumulative loss of radiative energy equals the gain in kinetic energy
of the wind material (Abbott \& Lucy 1985). 
We assume that each individual photon scatters coherently in the
co-moving frame. In the observer's frame, this implies that each
photon-ion interaction results in an exchange of momentum and energy
(except for the case that the incident and emerging photon happen to
have the same direction). As the absorptions and scatterings of photons 
in the wind depend on the wind density, this 
allows for a derivation of a globally consistent mass-loss rate -- for a pre-described flow structure. 
Our approach is described in detail in Vink et al. (1999).
We note that as the kinetic wind energy scales with
$\vinf^{2}$, the derived mass-loss rates depend on the choice of the
terminal flow velocity.
 
The chemical species that are explicitly calculated in non-LTE are H,
He, C, N, O and Si (see de Koter et al. (1997) for the details of the adopted atomic models). 
The iron-group elements, which are important for
the radiative driving and consequently for $\dot{M}$, are treated in a
generalized version of the ``modified nebular approximation'' (Lucy
1987, 2001, see also Schmutz 1991).  We include Fe ionization stages
up to {\sc ix}, and 
we note that the first 30 elements of the periodic table are all
included in the Monte Carlo computations for the line force. Although
the cumulative effect of all these species on the radiative
acceleration is properly included in our models, 
we will particularly highlight the roles of CNO, the dominant intermediate
mass elements, and Fe, the dominant heavy metal.

As we wish to investigate the effect of a different chemical
environment on the mass-loss rate of WR stars, we
exclude the self-enriched element of carbon in WC stars, from what we
call ``$Z$''. Therefore, where we address the
\mdot\ -- $Z$ dependence in WC stars, we mean a scaling with for
instance the Fe abundance, realizing that the overall WR metal content
is much larger (because of the large amounts of carbon produced by the
WC stars themselves).  Where we vary the Fe abundance, we also scale
the other metal species (S, Cl, Ar, etc.) in accordance to their solar
values. In the WN case, we assume that the CNO-cycle affects
the relative abundances of C, N, and O in amounts that are similar to 
our study for Luminous Blue Variables (LBVs; Vink \& de Koter 2002 and references therein), 
so that the total sum of C+N+O abundance is constant, and scales with the iron abundance.

We note that the stellar and ISM abundances are equal for elements
such as Fe, but the stellar and ISM carbon abundance can be vastly
different.
It is therefore important to be very careful with our
definition of $Z$. We also note that extragalactic metallicities are
usually derived from nebular oxygen lines.  The ISM oxygen and
Fe abundances do not necessarily scale with the solar
values, due to the differing roles of SNIa, which release Fe, and SN
II, which release large amounts of $\alpha$ elements, but less Fe. [$\alpha$/Fe]  
scalings can be adjusted using Table~5 by Vink et al. (2001).  

The strengths of our approach are the natural way in which multiple
line and continuum interactions are accounted for, and the consistent
computation of an ionized stratification throughout the wind. We 
do however make assumptions, which are listed below.

\subsection{The wind dynamics} 
\label{sec:winddynamics}

The most important assumption in our modelling is that we calculate
$\dot{M}$ with a pre-described velocity law. To be more precise, we
adopt a $\beta$-type velocity law for the accelerating part of the
wind:
\begin{equation}
\label{eq_betalaw}
v(r)~=~\vinf~\left(1~-~\frac{R_*}{r}\right)^\beta
\end{equation}
Below the sonic point, a smooth transition from this velocity
structure is made to a velocity that follows from the photospheric
density structure. Vink et al. (2000) have shown that for O stars the
predicted mass-loss rate is insensitive to the adopted value of
$\beta$ in the range of $\beta$ = 0.7 -- 1.5.  Therefore, we routinely
apply a value of $\beta$ = 1 for O stars, which is consistent with the
spectral line modelling of these stars (e.g. Puls et al. 1996), as
well as with modified CAK theory, where $\beta$ $\simeq$ 0.8
(Pauldrach et al. 1986).  There are observational indications that
$\beta$ may be significantly larger than 1 for denser winds, at least
in the outer regions (e.g. Barlow \& Cohen 1977, Antokhin \&
Cherepaschuk 2001), although this is still under debate (e.g. Ignace
et al. 2003).  On numerical grounds, Hillier \& Miller (1999) proposed
a ``double'' $\beta$ law, with $\beta_{1} = 1$ for the inner and
$\beta_2 = 50$ for the outer wind, which appears to be a good
approximation of the self-consistently derived velocity law for the WC5 star
modelled by Gr\"{a}fener \& Hamann (2005).  Here we do not apply a
2-step velocity law, but we do explore the difference between a
``low'' value of $\beta$ ($\beta$ = 1), and a larger value 
($\beta$ = 3) as to test the sensitivity of the velocity law
assumption, and to investigate whether our choice of $\beta$
influences the derived WR wind metallicity dependence
significantly. We find that this is not the case (Sect.~\ref{s_beta}).

We stress that although our approach allows for a derivation of the
mass loss that can be driven by the photons passing through the
outer atmosphere, it does not treat the
initiation and driving of the wind in the deepest layers at very large
optical depth (above a Rosseland optical depth of 30). 
To do so properly would require (i) taking account of the
back-warming of Fe at large depths (as this affects the
temperature and ionization structure), (ii) a co-moving frame
treatment of the lines (we adopt an improved Sobolev description
appropriate for regions higher up), as well as (iii) accounting 
for ionization stages above stage {\sc ix} (Gr\"{a}fener \& Hamann account for stages up to
Fe~{\sc xvii}). Although one should be aware of this, we 
again refer to the modest differences in overall wind properties reported 
by Gr\"{a}fener \& Hamann between the case where the wind dynamics was solved for, 
and the case for which a velocity stratification was adopted.

\subsection{Sphericity}

We assume that WR winds are spherically symmetric. Given the absence
of line depolarisations in 80\% of the WR stars analysed with the technique of 
linear spectropolarimetry (Harries et al. 1998), sphericity is
anticipated to be an appropriate assumption at Galactic $Z$ (Note however that
rotation may become more important at lower $Z$; see Sect.~\ref{s_impl}).

\subsection{Line branching}

In our modelling of the radiative force, all line interactions are assumed to be
scatterings. Lucy (2002, 2003) has developed techniques that can
accommodate for this potential deficiency, and Sim (2004) has investigated the
effects of line branching on Monte Carlo mass-loss predictions, and 
found that neglecting line branching can overestimate \mdot\ by up
to a factor of three for WR stars. Sim's method however does not
account for continuum absorption and subsequent thermal emission, which are 
included in our model. This implies that we have
a channel to redistribute photons towards longer wavelength, away from
the flux maximum, so that we cannot directly compare to
these findings. Nonetheless, we anticipate that the neglect of line
branching could cause an overprediction of the mass-loss rate by a factor of 
1--3 at galactic $Z$. As the importance of line branching is expected to decrease at
lower $Z$, we anticipate that by neglecting line branching we may 
slightly overestimate the exponent in the $\mdot$--$Z$ relation.

\subsection{Wind clumping}
\label{sec:nowindclumping}

It has been established that small-scale clumping of the
outflowing gas has a pronounced effect on the ionization structure of
WR atmospheres, as it favours recombination over ionization 
processes (e.g. Hillier 1991). In addition, clumping may have a 
direct effect on the radiative driving, although these effects 
may be complex and challenging to predict (Schmutz 1997, Hillier 2003). 
In more quantitative terms, the observed emission line spectrum -- formed
through recombination -- is essentially invariant for the ratio
$\mdot/\sqrt{f}$, where $f$ is the volume clumping factor describing
the mean density in the flow $\overline{\rho} = f \times \rho$,
where $\rho$ is the density in the clumps. The interclump medium is
assumed to be void. 
Typical values for the clumping factor in WR
winds are found to be 0.1 -- 0.25 (Hillier \& Miller 1999, Hamann \& Koesterke 1998).  
The ionization structure of a clumped wind will behave similarly to that of an unclumped 
wind with an $\sqrt{f}$ times lower mass-loss rate.
There may be a dependence of the clumping factor on
metallicity. If it holds true that weaker winds (e.g. O star winds) are less clumped 
than WR winds  (e.g. Lamers \& Leitherer 1993, but see Bouret et al. 2005), one might speculate 
that the winds of WR stars at lower $Z$ may also be less clumped than those of Galactic WR stars.

Nevertheless, we emphasize that the neglect of clumping could effect the results 
presented in this paper.

\subsection{Ion decoupling}

In line-driven winds, the ions that are most effective in absorbing
photons (i.e. those with many lines in the region of the flux maximum
of the star; notably C, N, O, and Fe) redistribute the momentum they
have gained -- through friction -- to ions that are only marginally
effective line absorbers.  These latter more ``passive ions''
include H and He, containing most of the mass. If
the wind density becomes lower, the coupling between active and
passive ions may disappear, potentially giving rise to multicomponent
effects (e.g. Lucy \& Solomon 1970; Castor et al. 1976). In O stars of
Galactic metallicity, this ion-decoupling will only occur in the
outermost regions of the wind, where the flow reaches its asymptotic
velocity -- and the mass-loss rate is not affected.

For the lower density winds at extremely low $Z$, ion-decoupling may
become important for the mass-loss rate.  It has been
proposed that runaway instabilities may occur in a multi-component
wind (Owocki \& Puls 2002, Krti\v{c}ka \& Kub\'{a}t 2002), and
possibly even fallback of the passive ions. Krti\v{c}ka et al. (2003)
present a relation defining the mass loss -- metallicity domain
where multicomponent effects become relevant. Note that their
diagram is intended for normal O stars. Ignoring this fact, and
applying this diagram for the WR model parameters discussed in our
study indicates that multicomponent effects start to play a role below
a metal content of $Z/\zsun$ $\sim 1/300$. However, two remarks are in order:
first, at low $Z$ the relative contribution of hydrogen and helium to
the line force will increase (see Vink et al. 2001), therefore
these elements become ``less passive''. At $Z = 1/300 \zsun$ this
contribution may amount to several tens of percents ($\sim$ 10--30 \%). 
Second, and especially relevant for WC stars, the mass fraction
contained in ``passive ions'' decreases with an increasing surface
abundance of carbon and oxygen. For the WC model that we study here,
already half of the mass in contained in carbon. Therefore, if
ion-decoupling would occur, one may expect that the predictions
presented in this paper overpredict \mdot\ by at most a factor of
two.


\section{The WR parameters}
\label{s_parameters}

This study is not intended to be a full parameter study of
WR stars, but is deliberately limited to very specific choices of
stellar properties. For an accurate modelling of higher temperatures,
i.e. earlier spectral types, the line list is required to be complete
for the higher ionization stages.  As we are principally interested in
the metallicity dependence, we choose to model ``late'' WR stars of
both flavours: WNL, the late-type nitrogen-rich stars, and WCL, the
late-type carbon-rich ones. Our strategy is as follows: we choose
typical WN/WC stellar parameters such as luminosity and effective
temperature, based on studies using the most sophisticated
``standard'' WR star models of Hillier \& Miller (1998) and 
Gr\"{a}fener et al. (2002), as
well as characteristic abundances, known from spectral synthesis
modelling, and evolutionary models. 
We then vary the metal abundance $Z$, such that we may study the
effects of a different host galaxy environment on \mdot.

In reality, both stellar structure and evolution are expected to
be $Z$ dependent as well, however we do not take this into account at
this stage (but see Sect.~\ref{s_impl}).  The crucial point here is
that if~ Fe dominates the mass loss of WR stars, mass loss is
anticipated to be $Z$-dependent, and the precise chemical enrichment
will be less relevant.  However, if it turns out that lighter
elements, such as the intermediate mass element carbon set the mass
loss for WR stars, mass loss is not expected to be dependent on the
chemical environment after all.

Although we do not attempt to reproduce a most realistic WR wind model
{\em per se}, we do aim to construct a representative model.  
This is no trivial task for the optically thick winds of WR stars.  In
particular, the effective temperatures of WR stars are not well
defined, since the radius is strongly wavelength dependent (e.g. Schmutz 1991). Nugis \& Lamers (2000, 2002) argue that the only
meaningful temperature, i.e. one that is independent on the wind strength, is
the one that corresponds to the hydrostatic core, as computed in
stellar evolution codes. Hence, Nugis \& Lamers (2000) list
effective core temperatures, termed $T_{\rm evol}$, well in excess of 100~000 K.
Their luminosities are of the order of respectively \logl\ $\simeq$ 5.8 for WN
stars, and \logl\ $\simeq$ 5.3 for WC stars.

Due to the large geometrical extent of Wolf-Rayet atmospheres
the effective temperature cannot be unambiguously defined. Typically,
two effective temperatures play a role, see e.g. Dessart et al. (2000)
for WC stars, and Herald et al. (2001) for WN stars. The first is based on 
the total flux at the inner boundary radius \rstar\ of the model, and 
is used to define the stellar luminosity $L_{*} = 4 \pi R_{*}^2
\sigma T_{*}^4$. 
We note that recent studies incorporating a
self-consistent treatment of line blanketing, such 
as for instance that of Herald et al. (2001) for the WN8
star WR\,40, tend towards smaller core radii and larger \tstar. This has 
led to a present-day luminosity of \logl = 5.6 for this object.
The second temperature that is frequently used is the effective 
temperature at the radius where the Rosseland optical 
depth equals 2/3.  
In our studies, we define the stellar radius \reff\ and effective 
temperature \teff\ slightly differently, namely at 
the point where the thermalization optical depth measured in the center 
of the photometric
$V$ band (at 5555 \AA) equals $1/\sqrt{3}$ (see Schmutz et al. 1990
and de Koter et al. 1996 for detailed discussions). 
We note that these two definitions are only marginally different.

If the mass-loss rate for Wolf-Rayet stars is a function of metal
abundance, the stellar radius may be expected to depend on $Z$ as well,
because it is the wind strength that determines the location of \reff.  
This adds a complication to the investigation of the mass-loss
dependence of Wolf-Rayet stars. Recent WC spectral analyses including 
line-blanketing show values for \tstar\ in the range 57 -- 110 kK (Dessart et al. 2000, Crowther 
et al. 2002, Gr\"{a}fener et al. 2002) -- i.e. in some case well below the expected $T_{\rm evol}$ -- 
and large geometrical extensions of WC atmospheres. If we adopt a
constant \tstar\ and $R_{*}$ for all $Z$ models, we may
find that we are comparing models with widely different \teff.  If, on
the other hand, we adopt a constant \teff\ (by adjusting \rstar) for
all models, we might ignore the fact that at low metallicity such
relatively cool stars may not actually exist.

In this study, we opt to keep \tstar\ constant for both the WN
and WC stars. We find that although the predicted mass-loss rates for
Galactic metal content are of order $10^{-5}$ \msunyr, this does not
lead to the formation of a truly extended optically thick wind.
Therefore, in our computations the effective temperature will not increase 
strongly if we lower the metal abundance. We do not claim that the
effective stellar radius is not affected by a change in metal
content in real WR stars; it merely reflects the fact that an adjustment 
of the star in the sub-photospheric region due to a change in $Z$
is not treated in our models.

\subsection{The WN and WC model parameters}

For our typical WN star, we choose the following stellar parameters
(see Table~\ref{t_params}): \tstar\ = 40\,000 K, and \logl\ = 5.62.
The stellar mass of Wolf-Rayet stars cannot be obtained from 
spectral synthesis modelling, and we have therefore adopted a value 
of $M$ = 20 \msun\
based on the compilation of Nugis \& Lamers (2000) for late WN stars.
The terminal wind velocity we adopt is \vinf\ = 840 \kms, as found by
Herald et al. (2001) for the WN8 star WR\,40. The abundances were taken
from Table 9 in Herald et al. (2001).  Note the relatively high
nitrogen abundance, as typical for WN stars of course, and the high
helium abundance, $Y$ = 0.83 (by mass). As helium is not {\it directly} relevant 
for the line driving at galactic metallicities (although this changes at
lower $Z$; Vink 2000), it may well be the case that the mass-loss metallicity dependence for these 
stars is similar to that of O stars.

\begin{table}
\caption[]{Adopted parameters for our typical Galactic WN and WC stars. 
           The Galactic WN model is roughly representative of the WN8 
           star WR\,40; the typical WC model resembles the WC8 star WR\,135. The relevant 
           abundances are given by mass ratio.}
\label{t_params}
\begin{tabular}{ccc}

\hline\noalign{\smallskip}

Parameter   & WN          &  WC         \\
\hline
\tstar\ (K)  & 40000       &  40000     \\
\logl\       & 5.62        &  5.36        \\
$M$ (\msun)  & 20          &  10          \\
\vinf\ (\kms)& 840         &  1200    \\
X            & 0.15        &  --     \\
Y            & 0.83         &  0.5          \\ 
C            & --           &  0.5           \\

\hline
log \mdot(\msunyr)      & $-$4.89      &  $-$5.34       \\
\noalign{\smallskip}
\hline
\end{tabular}
\end{table}


We first note that WC stars are believed to be He burning stars, theoretically 
expected to lie on the He-main sequence with large surface temperatures (in excess of 
100~000 K). Observationally, significantly lower effective temperatures are seen, presumably due to large amounts 
of mass loss at Galactic metallicity. We have chosen the
following stellar parameters for our WC model: \tstar\ = 40\,000 K, and \logl\ = 5.36. 
The adopted mass is $M$ = 10 \msun, again based on the compilation of
Nugis \& Lamers (2000) for late WC stars, and in line with the
expectation that a later evolutionary phase results in a smaller mass,
as a result of mass loss.  The terminal wind velocity we adopt is
\vinf\ = 1200 \kms, as found by Dessart et al. (2000) for the WC8 star
WR\,135. Nugis \& Lamers (2000) found $N_{\rm C}/N_{\rm O}$ $\simeq$ 5 for all WC subtypes, but
a $N_{\rm C}/N_{\rm He}$ that is spectral-type dependent,
ranging between 0.18 -- 0.36. These abundance
ratios are by number.  Based on the above, it is clear that oxygen is
not expected to be an important line driver, however carbon should
become crucial at very low iron abundance.  
We adopt a ratio $N_{\rm C}/N_{\rm He}$ of 1/3, which corresponds
to an equal mass contribution of these two elements. Such a high ratio
maximizes the potential effect of carbon line driving, compared to
that of Fe. (If we would take a lower ratio, carbon would start to
dominate at a lower $Z$ value.)

Note that all these parameters can be adjusted to obtain a better
agreement between our mass-loss predictions and the currently favoured
observed rates. In particular, an increase in the carbon over helium
ratio, a decrease in the stellar mass, an increase in the star's
luminosity, a decrease in its terminal flow velocity, or a more rapid
wind acceleration, may all cause an increase of the mass-loss rate.
The purpose of this study is, however, not to establish the optimum
conditions for wind driving, but to create WR models with reasonable
parameters, and to investigate the $Z$ effect on the mass-loss rate.

\section{The predicted mass-loss rates}
\label{s_massloss}

\subsection{The typical WN model}

We have used the method by Lucy \& Abbott (1993), as shown in their
Fig.~1, to search for the mass-loss rate. This mass-loss determination
method yields rates that are globally consistent, and has been used 
for OB supergiants (Vink et al. 1999) and LBVs 
(Vink \& de Koter 2002).
Our typical WN model yields a mass-loss rate of log \mdot (\msunyr) $= -4.89$.
This number is in fair agreement with observations.
Clumping-corrected mass-loss rates for WN stars give values in
the range $\log \mdot = -4.4$ to $-5.0$ (e.g. Nugis \& Lamers 2000,
Hamann \& Koesterke 2000). The predicted value is thus representative
for the lower range of observed mass loss (i.e corresponding to 
weak lined WN stars).

\begin{figure}
\centerline{\psfig{file=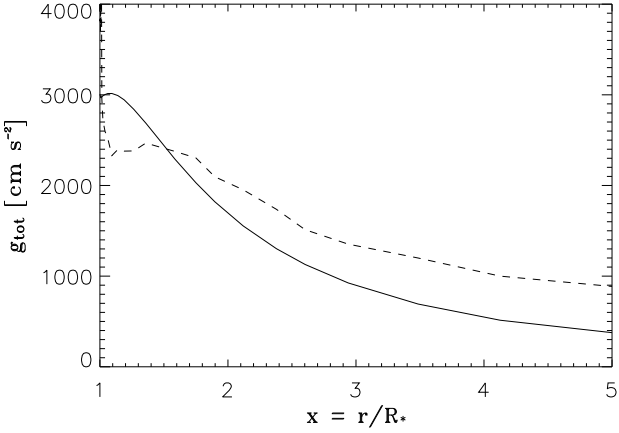, width = 9 cm}}
\caption{The computed radiative acceleration (dashed line), and the
         one implied by the adopted $\beta$ = 1 velocity law
         (solid line) versus radius -- for the Galactic WN model. The
         cross-over point, i.e. where the model changes from an
         underprediction in the inner wind to an overprediction of the
         line acceleration in the outer wind, occurs at about 1.5
         \rstar.}
\label{f_gradcomp_wnl}
\end{figure}

As we have adopted a wind velocity law, and only obtained a
globally consistent mass-loss rate, the system is not necessarily
dynamically consistent. Although we do not aim to derive a model
that is locally consistent in the transfer of bulk momentum from the
photons to the gas, we may still test the quality of our
velocity law by comparing the radiative acceleration that we compute
to that implied by the adopted velocity law via the equation
of motion.  Figure~\ref{f_gradcomp_wnl} shows the computed radiative
acceleration by the dashed line, and the one imposed by the velocity
law by the solid line. Both show a similar qualitative behaviour
with radius, but there are systematic differences. Our models
underpredict the line force in the inner wind, and overpredict it in
the outer wind. This is similar to what was found by Lucy \&
Abbott (1993) and Schmutz (1997).  The likely reasons for these
discrepancies have been listed in Sect.~\ref{sec:winddynamics}
(i.e. no back-warming due to Fe; no iron ionization stages above 
{\sc ix}, no co-moving frame treatment). We restate
that Gr\"{a}fener \& Hamann (2005) found only a modest difference
between the predicted mass loss based on a self-consistent treatment
and the one found using a pre-described velocity structure.

\subsection{The mass-loss metallicity dependence for WN winds}
\label{s_wnlz}

\begin{figure}
\centerline{\psfig{file=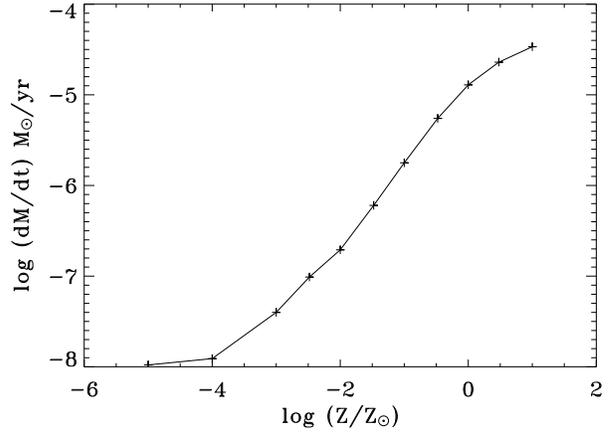, width = 9 cm}}
\caption{The $\mdot(Z)$ dependence for WN stars for the case
         $\beta = 1$. The relation shows
         a power-law behaviour in the range $10^{-3} \lesssim Z/\zsun
         \lesssim 1$ with an index $m = 0.86$. At higher/lower metal
         abundances, the relation appears flatter (see text for explanations).}  
\label{f_mdotz_wnl}
\end{figure}

Using our WN model as a starting point, we systematically vary
the metallicity. The results are given in the second 
column of Table~\ref{t_massloss}, and are plotted in
Fig.~\ref{f_mdotz_wnl}.  The figure shows an \mdot\ versus $Z$
dependence that is more or less linear between $Z/\zsun =$~$10^{-3}$
-- 1. It flattens off for supersolar values of $Z$,
because the iron lines -- present
throughout the spectral region of the bulk 
flux -- become saturated. Phrased differently, the spectrum
``runs out of continuum windows between saturated spectral lines''
through which photons could escape (see Vink et al. 2001 who describe
the effect for OB stars).  The mass loss -- metallicity dependence 
also flattens off for the lowest $Z$ models at $Z/\zsun \la 10^{-3}$. 
The reason is also similar to that of OB stars: CNO, and to some extent even hydrogen and helium
(especially at extremely low $Z$), take over the role that Fe
plays in the line driving at Galactic $Z$ (Vink et
al. 1999, Puls et al. 2000).  
Over the range $Z/\zsun =$~$10^{-3}$ -- 1, the mass
loss vs. metal content dependence behaves as a power-law, i.e.
\begin{equation}
   \log \dot{M} \propto m\,\log(Z/\zsun),
   \label{eq:mdotz}
\end{equation}
with index $m = 0.86$.
The power-law dependence 
for the WN star is similar to that of OB stars, as the
line driving in both cases is primarily due to Fe lines. 
We note
that this result is based on only one set of stellar parameters. Also,
the fitting errors may be smaller than systematic errors associated
with our modelling.

\begin{table}
\caption[]{Predicted WR mass-loss rates for different metallicities
           for both WN and WC stars. The Galactic WN model is 
           roughly representative of the WN8 star WR\,40, the typical
           WC model resembles the WC8 star WR\,135}
\label{t_massloss}
\begin{tabular}{ccccc}

\hline\noalign{\smallskip}
$Z/\zsun$   & WN          &             &  WC          &    \\
            & $\beta =1$   & $\beta =3$  &  $\beta =1$   & $\beta =3$\\
\hline
10          & $-$4.47      & $-$4.68     & $-$4.93       & $-$5.05\\
3           & $-$4.64      & $-$4.82     & $-$5.14       & $-$5.30\\
\hline
1           & $-$4.89      & $-$5.11     & $-$5.34       & $-$5.50\\
\hline
$1/3$       & $-$5.26      & $-$5.49     & $-$5.62       & $-$5.77\\
$1/10$      & $-$5.75      & $-$5.94     & $-$5.99       & $-$6.11\\
$1/33$      & $-$6.22      & $-$6.44     & $-$6.22       & $-$6.34   \\
$1/100$     & $-$6.71      & $-$6.85     & $-$6.39       & $-$6.52   \\
$1/333$     & $-$7.01      & $-$7.16     & $-$6.57       & $-$6.67   \\
$10^{-3}$   & $-$7.40      & $-$7.40     & $-$6.74       & $-$6.77   \\
$10^{-4}$   & $-$7.91      & $-$7.62     & $-$6.83       & $-$6.85   \\
$10^{-5}$   & $-$7.98      & $-$7.67     & $-$6.85       & $-$6.86   \\

\noalign{\smallskip}
\hline
\end{tabular}
\end{table}

Our mass-loss computations have been performed with a constant wind
velocity, and thus, the $Z$ dependence actually applies to the wind
{\it momentum}, $\mdot \vinf$.  In reality, \vinf\ may also be
$Z$-dependent, which would imply that the power-law index $m =
0.86$ could be an upper limit.  There are no CAK-type scaling
relations available for WR winds.  Observationally, the dependence of
\vinf\ on $Z$ is also unknown.  For OB stars, index values $p$
for a $\vinf(Z)$ dependence given by $\vinf = Z^{p}$ of the order of
0.10 -- 0.20 have been noted (e.g. Garmany \& Conti 1985, Prinja 1987,
Walborn et al. 1995). Evans et al. (2004) have recently
reported that there appears to be no significant difference in \vinf\
between B supergiants in the Galaxy and in the Magellanic
Clouds. So, even for OB stars there is no
clear-cut observational evidence for a $\vinf(Z)$ dependence. And if
there is one, it is likely only a modest one.

Turning from observations to expectations, one would perhaps
anticipate there to be a $\vinf(Z)$ dependence, as the global line
force is lower at lower $Z$, resulting in lower \mdot\ and lower
\vinf.  Also in CAK theory the value of the force multiplier $\alpha$
is lower, hence \vinf\ is expected to be lower (Kudritzki \& Puls
2000, Puls et al. 2000).  Therefore, OB wind theory predicts values of
$p$ of the order of 0.13 (e.g. Leitherer 1992).  However, if the mass-loss rate 
is indeed lower at low metallicity there may be a {\it second}, less well-documented, 
effect that operates in the opposite direction.  To some extent, the line 
driving is expected to {\it increase} in the outer wind -- yielding a 
{\it larger} \vinf -- as a result of the line acceleration being 
inversely proportional to the density in case of optically thick lines 
(at very low $Z$ optically thick lines due to CNO may gain in relative
importance over the contribution of optically thin iron
lines.) 
At Galactic $Z$, this second argument is also at work: it operates in B1 supergiants 
in relation to the bi-stability jump (Vink et al. 1999), as well as for the weak 
winds of B main-sequence stars, for which Babel (1996) finds particularly high terminal wind velocities.

It is clear that the \vinf\ -- $Z$ dependence
is still relatively unexplored territory requiring future study, both
observationally and theoretically.  We conclude that the
power-law indices $m$ we derive for the mass-loss rate -- at constant
\vinf -- are likely upper limits, but that the actual values 
are expected to be rather close to these numbers.

\subsection{The typical WC model}

We now turn to carbon-rich stars.  The predicted mass-loss rate
for our typical WC model is found to be $\log \mdot = -5.34$.  A
recent analysis of WC stars by Dessart et al. (2000) report values in
the range $ -4.5 \lesssim \log \mdot \lesssim -5.1$. Crowther et
al. (2002) find for a WC star of comparable luminosity a mass loss
$\log \mdot = -4.7$ and a wind clumping factor $f = 0.1$. So, we
probably underpredict the mass loss by about a factor two to four.

\begin{figure}
\centerline{\psfig{file=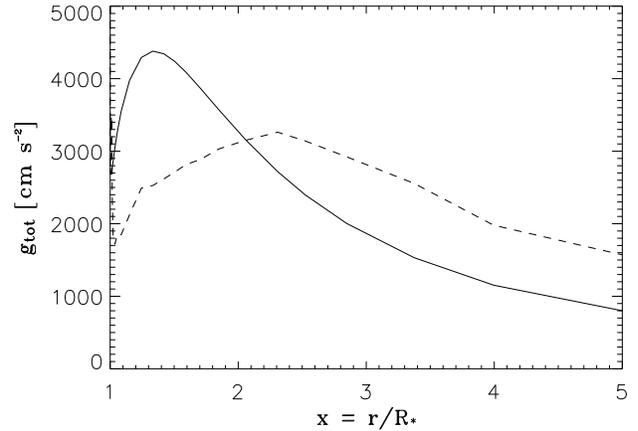, width = 9 cm}}
\caption{The computed radiative acceleration (dashed line), and the
         one imposed by the adopted $\beta$ = 1 velocity law (solid
         line) versus the wind radius -- for the Galactic WC
         model. Note that the cross-over point, where the model
         changes from an underprediction in the inner wind to an
         overprediction of the line acceleration in the outer wind,
         occurs at about 2 \rstar.
         }
\label{f_gradcomp}
\end{figure}

We again test the quality of our velocity law by comparing the
radiative acceleration that we compute with the Monte Carlo model to
that implied from the adopted velocity law via the equation of motion.
Figure~\ref{f_gradcomp} shows the computed radiative acceleration by
the dashed line, and that imposed by the velocity law by
the solid line. Both show a similar qualitative behaviour with
radius, but we underpredict the line force in the inner wind, and
overpredict it in the outer wind.

\subsection{The mass-loss metallicity dependence for WC winds}
\label{s_wclz}

\begin{figure}
\centerline{\psfig{file=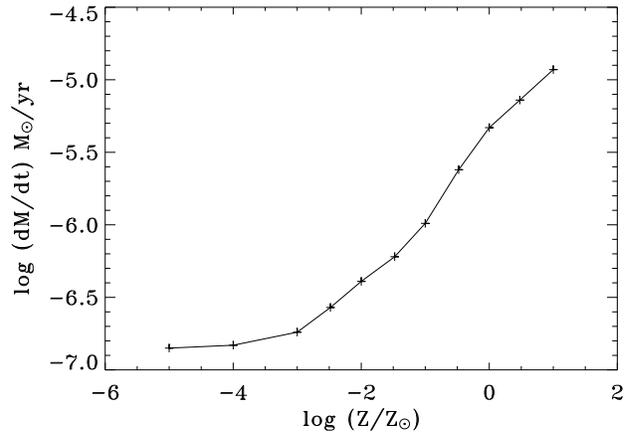, width = 9 cm}}
\caption{The $\dot{M}(Z)$ dependence for WC stars for the case
    $\beta = 1$. The relationship can be devided into four linear
    segments. Describing the mass loss vs. metallicity relation as a
    power-law, i.e. $\log \mdot \propto m \log Z$, we find $m = 0.40$
    for $(Z/\zsun) \gtrsim 1$; $m = 0.66$ 
    for $10^{-1} \lesssim (Z/\zsun) \lesssim 1$; $m = 0.35$ 
    for $10^{-3} \lesssim (Z/\zsun) \lesssim 10^{-1}$. Note that at 
    $Z \lesssim 10^{-3} \,\zsun$ the mass loss becomes roughly constant.
    At such low metal content the wind driving is dominated by
    radiation pressure on carbon lines.
    }
\label{f_mdotz}
\end{figure}

Keeping all stellar parameters fixed we now systematically vary
the metal content of the WC star. The results are given in the third
column of
Table~\ref{t_massloss}. Figure~\ref{f_mdotz} shows the \mdot\ versus
$Z$ dependence for $\beta = 1$. The dependence is more complex
than in the case of WN stars, as several linear parts in this
log--log plot can be identified. The linear part with the steepest
slope occurs at $Z/\zsun = 1/10 - 1$. The power-law index of this part of
the $\mdot(Z)$ function (see Eq.~\ref{eq:mdotz}) is $m = 0.66$.

At higher metal content the dependence flattens ($m = 0.40$), caused
by the same line saturation effect as described in Sect.~\ref{s_wnlz}
for WN stars. At metal abundances below 1/10 of the solar value, the
dependence also flattens ($m = 0.35$), relative to the behaviour at Small
Magellanic Cloud to Galactic abundances. A further flattening occurs
at $Z \simeq 10^{-3}$ -- essentially forming a plateau where mass
loss no longer decreases with $Z$. Even though the luminosity of
the WC stars is about a factor two less than that for the WN star,
the predicted mass-loss rates in this very low $Z$ regime is more than
an order of magnitude larger. The reason for the tendency towards
constant mass loss at $Z \simeq 10^{-3}$ is the ever increasing
dominance of driving by spectral lines of carbon, responsible for
almost half of the mass in the atmosphere.

\subsection{Varying the $\beta$ exponent of the velocity law}
\label{s_beta}

\begin{figure*}
\centerline{\psfig{file=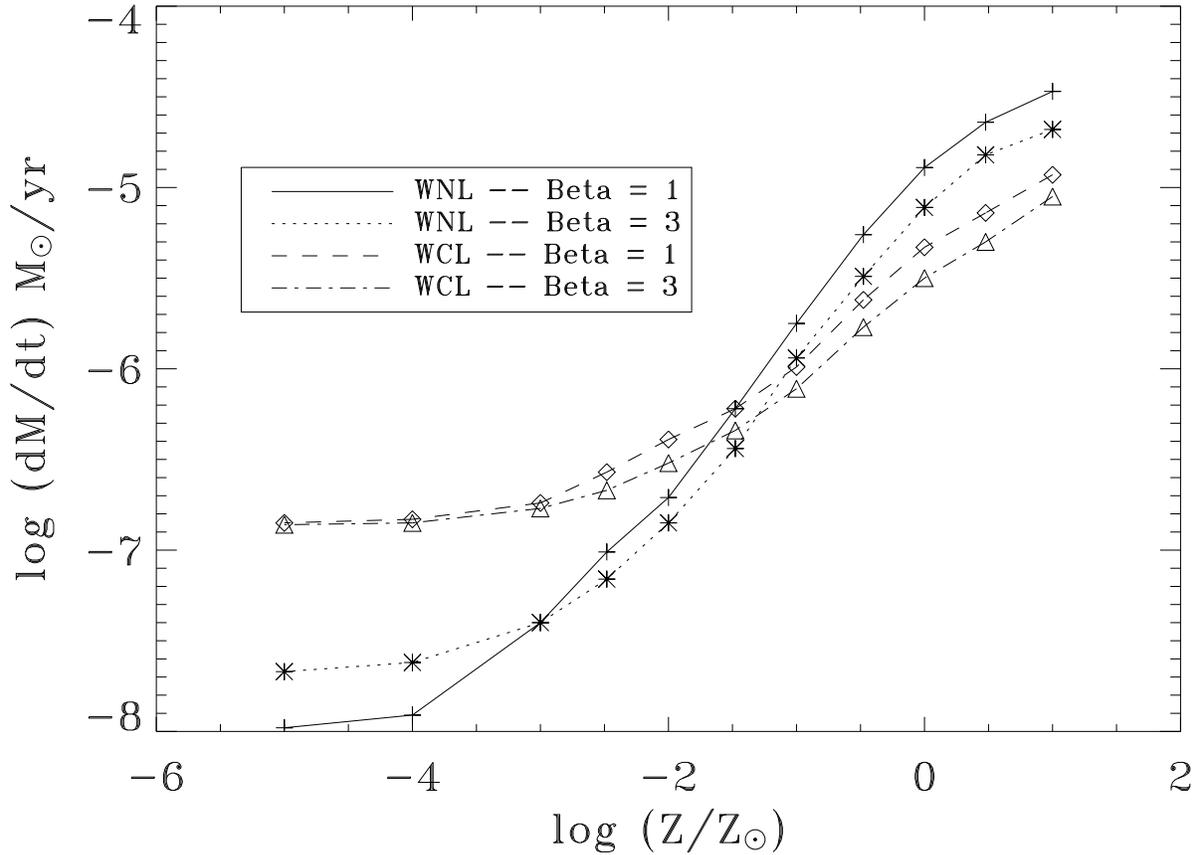, width = 18.0cm}}
\caption{The Mass-loss versus $Z$ dependencies for both our WN and
         WC models, for velocity law exponent parameters $\beta$ = 1
         and 3. The crosses are for WN stars with $\beta$ = 1, the
         stars for WN and $\beta$ = 3. The open diamonds represent
         WC stars with $\beta$ = 1, and the open triangles indicate
         WC models with $\beta$ = 3. The predictions, both in
         terms of the absolute mass-loss as well as in the slope $m$
         of the $\mdot(Z)$ relation do not depend strongly on the
         adopted value of $\beta$.}
\label{f_mdot_wr_all}
\end{figure*}

So far we have discussed WN and WC models with
a $\beta$~=~1 velocity law.  
As mentioned
before, there are indications that the velocity exponent may in fact
be less steep for the outer wind, whilst $\beta$ = 1 may be a proper
representation for the inner wind.
We wish to explore whether our choice of $\beta$ influences the
derived WR wind metallicity dependencies. We have therefore repeated
the computations of Sects~\ref{s_wnlz} and \ref{s_wclz} for a
$\beta$~=~3 velocity law.  The results are given in the third and
fifth column of Table~\ref{t_massloss}, and are plotted in
Fig.~\ref{f_mdot_wr_all}.  Although there are notable changes in the results
for both the WN and WC stars between the $\beta$ = 1 and $\beta$ = 3
velocity exponents, the general shapes are very similar, and
the differences in log \mdot\ of the individual models are only of the
order of 0.1 -- 0.2 dex or less. 
Differences in the slope $m$ of
the mass loss -- metallicity relations amount to only 0.01 -- 0.04 dex. 
This suggests that our results are not too sensitive to the adopted
velocity law which, while reassuring, it does not remove the
need for future hydrodynamic modelling.


\section{Discussion and Implications}
\label{s_impl}

Before we consider the implications of our metallicity dependent
mass-loss rates for WR stars, we first relate the mass-loss rates
derived for WR stars in this study to previous results for O
supergiants (Vink et al. 2000). We compare the Galactic WN and WC
stars to O stars having the same luminosity and effective temperature.
The mass of the O stars is derived from the evolutionary tracks of
Meynet et al. (1994), yielding 43.5 and 35.4 \msun\ for \logl\ = 5.62
and 5.36 respectively. We find mass-loss rates $\log \mdot \,(\msunyr) =
-5.73$ and $-6.14$ for the O-type counterparts of the WN and WC
stars, i.e. the mass-loss rates of the WR stars are a factor of
$6 - 7$ higher.
Our WR model properties differ in essentially three aspects relative 
to O stars: they have a much lower mass; they have a
different chemical composition; and their terminal flow velocities are
reduced. At a Galactic metal content, the main effect causing the
difference is the reduced mass (see also Vink \& de Koter 2002). 
The results presented in this paper show that chemical processing effects become important 
at very low metallicity.

At low $Z$, the Wolf-Rayet models show a flattening of the $\mdot(Z)$
relation. The limiting (minimum) mass-loss rate for the WC model is
$\sim 10^{-7} \msunyr$; for the WN model it is an order of magnitude
less. Given that for the WC model already half of the mass resides in
carbon, we do not expect the minimum $Z$ at which our $\mdot(Z)$
relation flattens to be significantly different in the case of WO
stars, although these objects have significantly larger terminal velocities than we have modelled. 
Note that the predicted mass loss in the WC phase is not
typical for only one initial mass, but for all $M_{\rm init} \gtrsim 40
\msun$, as Wolf-Rayet properties in the WC phase are similar for
all massive stars (see e.g. Maeder \& Meynet 1994).

Given that the lifetimes of massive stars are of order $10^{7}$ yr
and that mass loss tends
to increase with stellar age, we anticipate that the total mass that
is lost by a massive star at $Z \lesssim 10^{-3}$ due to a continuous
outflow caused by radiation pressure on spectral lines is only of the 
order of one \msun. 

This result immediately leads to the question whether WR stars will
actually form at very low $Z$, as the formation of a He-rich star requires
the removal of its H-rich envelope. It has sometimes been assumed that
this may not occur at all within the context of single-star evolution,
due to a lower mass loss at lower $Z$, and that the formation of WR
stars at very low metal content always requires a (close) binary scenario.
Perhaps surprisingly, searches for, and statistics of binary
WR stars in the low
$Z$ environments of the Magellanic Clouds by Foellmi et al. (2003a,
2003b) seem to indicate a lack of binaries.
Note that so far we have discussed the formation of WR stars in an
evolutionary sense. The defining observational characteristic for
He-rich stars to be called WR stars is the dominance of {\em broad
emission lines} in the spectrum. This type of spectrum is the result
of line formation in a dense, trans-sonic outflow. The fact that we
find the enriched stars to have a lower wind strength at lower Z implies
that at metallicities lower than $Z/\zsun$ $\la$ $10^{-2}$ the stars
would not be expected to be spectroscopically identified as WR stars
(if the only mechanism contributing to the outflow of matter is a
stationary line-driven wind).

Additional mechanisms may contribute to the loss of mass from massive stars
at low metal content, such as stellar winds and mass ejections due to
super-Eddington luminosities, LBV-type mass
ejections, pulsations and/or rapid rotation, and one may speculate
that this may lead to the formation of He-rich WR stars, which e.g. may well 
be required to form at low $Z$ for the popular collapsor model (Woosley 1993) for 
long-duration GRBs to work.
These additional physical effects may play a role in either {\em causing mass
loss at low $Z$} or in {\it enhancing the mass loss by line driving}.
During the LBV phase massive stars are thought to experience violent
mass ejections (see e.g.  Humphreys \& Davidson 1994 for a
review). 
The mechanism causing these outbursts, in which 
up to tens
of solar masses may be ejected (Voors et al. 2000a, 2000b, Smith et al. 2003), is not
understood. Therefore, it is unknown whether this mechanism may also
occur at extremely low $Z$, and if so, what amounts of mass can be
expelled.
The possible effect of pulsations on mass loss of massive stars at
very low metallicity have been studied by Baraffe et al. (2001), and are 
found to be much weaker compared to those at solar metallicity.

In recent years, much effort has been directed towards understanding
the role of rotation in stellar evolution. The role of rotation for stellar winds 
may be considerable, as is for instance quite dramatically illustrated by the 
wind shape of $\eta$\,Carina (van Boekel et al. 2003), which may be caused by extra radiation pressure 
from the stellar pole due to gravity-darkening and stellar oblateness of 
a rotating star (e.g. Pelupessy et al. 2000, Dwardakas \& Owocki 2002).
The significance for massive star evolution at very low $Z$ is that
stellar rotation is anticipated to be much faster (e.g. Meynet \&
Maeder 2002), which may lead to both a higher mass-loss rate
(e.g. Friend \& Abbott 1986, Langer 1998) as well as more efficient 
mixing (e.g. Maeder et
al. 2004).

If our predictions of the absolute mass-loss rates in the WR
phase of very low $Z$ stars are taken at face value ($\mdot \sim 
10^{-7} \msunyr$), one may be inclined to conclude that the effect of
mass loss is too limited to significantly affect the evolution of
massive stars in the early Universe, and that winds would only modestly 
contribute to the input of energy, momentum, and products of
nucleosynthesis into the local ISM relative to supernovae.  However,
if mass loss by line driving occurs in conjunction with
one or more of the physical phenomena discussed above (such as
rotation of the star close to the Omega-Eddington limit), {\it it may
be critical in causing a much larger mass loss}. Recent numerical 
simulations suggest that stars in the early Universe at low $Z$ were 
more massive than in the present-day Universe (e.g. Bromm et al. 1999, Abel et al. 2002), 
so that these stars were more luminous (e.g. Schaerer 2002, Marigo et al. 2003), and closer to 
the Eddington limit. Therefore, our key finding
that the mass loss flattens off at low $Z$ may indicate a 
role for mass loss in determining the properties and evolution of massive
stars in the early Universe, contrary to assumptions made in many cosmological studies
where the neglect of mass loss is primarily based on CAK-type wind 
models for O stars (e.g. Salvaterra \& Ferrara 2003, Madau et al. 2004).

\section{Summary \& Conclusions}
\label{s_concl}

We have performed a pilot study of Monte Carlo mass loss computations
for late-type Wolf-Rayet (WR) stars as a function of metal abundance, over the
range between $10^{-5} \le (Z/\zsun) \le 10$. 
Although our predictions
take the effects of multiple scatterings and ionization
stratifications into account, we have made assumptions that might
affect the quantitative outcomes of our study, such as the exponents
of the \mdot\ -- $Z$ dependence, as well as the ranges over which
these exponents are valid.
We have found that the winds of
nitrogen-rich Wolf-Rayet stars are dominated by a myriad of Fe lines, so
that the dependence of mass loss on $Z$ is similar to that of OB
stars. We find that the wind momentum scales as $Z^{0.86}$ in the
range $10^{-2}$ $\leq$ $Z/\zsun$ $\leq$ 1.

For more evolved WR stars, where carbon is an abundant species, the
winds are still found to be dependent on the Fe abundance, and hence
they depend on the chemical environment of the host galaxy, but 
with a mass loss versus metallicity dependence that is less steep than 
that of their nitrogen-rich counterparts.  
At lower metallicities, this dependence however becomes much weaker, and 
it no longer declines anymore once the metal abundance drops below ($Z$/\zsun) $\simeq$ $10^{-3}$.  
To be more precise, we find that the wind momentum
scales as $Z^{0.66}$ for the range between 1/10 $\la$ $Z/\zsun$ $\la$
1. At lower $Z$, it becomes approximately $Z^{0.35}$, and mass loss no longer declines
below ($Z$/\zsun) $\simeq$
$10^{-3}$.

Our finding that WR mass loss is $Z$-dependent is a new one, with
important consequences for black hole formation and X-ray
population studies in external galaxies (e.g. Soria et al. 2005), as well 
as the prediction of ionizing radiation from young starburst clusters 
and galaxies. Finally, we note our finding that mass loss 
no longer declines once the metal abundance drops below ($Z$/\zsun) 
$\simeq$ $10^{-3}$, owing to an increased importance of
radiative driving by intermediate mass elements, such as carbon. 
In combination with rapid rotation and/or proximity to the 
Eddington limit -- likely to be relevant for massive Population {\sc iii} stars -- this effect 
may indicate a role for mass loss in the appearance and evolution of these objects, as well as a potential role 
for stellar winds in enriching the intergalactic medium of the early Universe.


\begin{acknowledgements}

We thank the referee, Goetz
Gr\"{a}fener, for constructive comments that have helped to improve
this paper. JSV is supported by PPARC.

\end{acknowledgements}

\end{document}